-----------------------------------------------------
\magnification=\magstep1

\hsize=6truein
\vsize=8.5truein
\hfuzz 2pt
\null 
\tenrm 
\def \i{{\rm i}}
\def \e{{\rm e}}
\def \um{{\textstyle {1\over2}}}
\def \uq{{\textstyle {1\over4}}}

\def \udp{{\textstyle {1\over2\pi}}}

{\centerline
{PAIR PRODUCTION }}
{\centerline
{IN A TIME DEPENDENT MAGNETIC FIELD}}
\vskip 2pc
{\centerline
{ Giorgio CALUCCI\footnote*{E-mail: giorgio@ts.infn.it}}}
\vskip 1pc
{\centerline
{\it Dipartimento di Fisica Teorica dell'Universit\`a di Trieste,
Trieste, I 34014 Italy}}
{\centerline
{\it INFN, Sezione di Trieste, Italy}}
\vskip 2pc
{\centerline {Abstract}}
\narrower
\midinsert
{ The production of electron-positron pairs in a time-dependent magnetic field
is estimated in the hypotheses that the magnetic field is uniform over
 large distances with respect to the pair localization and it
is so strong that the spacing of the Landau levels is larger than the rest
mass of the particles. This calculation is presented since it has been
suggested that extremely intense and varying magnetic fields may be found around
some astrophysical objects.}
\endinsert
\vfill
\eject
{\bf 1. Statement of the problem}
\vskip 1pc
 Standard Q.E.D. clearly states that a static electric field of sufficient
strength and extension can create pairs of opposite charges through a tunnel
effect [1,2], but the same kind of production is not foreseen for magnetic 
fields in the same conditions,
unless one would take into account the production of monopole pairs.
The situation changes if the external fields vary in time and the effect of the
field variation on the pair production rates has been discussed in detail in
connection with the possibility of detecting these effects in experiments in
optics [3], where the time dependence of the field is explicitly taken into
account, with an essential negative answer, at least for the real pair
production.\footnote*{ The effect of virtual pairs in external fields is a
different issue, which is under investigation also experimentally[4], but it 
will not be discussed here.} 
It may be possible that pairs are created in the presence of very large field
strengths with time variation which is not too small; since it has been proposed
 that situations of this kind could
take place in some phases of neutron star evolution[5,6], a
preliminary investigation on the pair production in a very strong and non
constant magnetic field is presented here. No detailed quantitative
calculation is performed, but some order of magnitude estimates are given
in order to show that the model calculation that will be presented can be well 
suited for the conditions referred to above.

These conditions correspond to a very intense magnetic, 
over $10^{10}T$, which remains constant over distances which are large in 
comparison with the dimensions involved in the electron production, during the 
time of variation the spatial uniformity is conserved, in this first 
investigation only a change of strength at constant direction is considered.
\par
 The rapidity of change
cannot be too small in order for the production process not to be entirely
negligible, but since it involves macroscopic objects it will be slow enough to
allow the adiabatic treatment of the production process[7]. 
For the ratio between the magnetic and the rest energy, defined as 
$\Re=(2\hbar eB)/(m_e c)^2$, we obtain for the chosen values of $B,\;\Re >4.5$,
which allows some computational simplifications that will be used. 
\vskip 1pc
{\bf {2. Determination of the production rate}}
\vskip 1pc
The production rate is calculated here in the following external field
configuration: the magnetic field grows at fixed
direction, this could happen locally, in the case of a
collapsing neutron star, when a compression of the magnetic field takes
place[6].
\par
In order to fix notations and conventions some well known features of the
relativistic electron in static uniform magnetic field are rewritten here.
\par 
The Dirac equation\footnote*{ here $e>0$ so the charge of the electron is $-e$.} 
$H\psi=[\vec \alpha \cdot (\vec p +e\vec A)+\beta m]\psi=\epsilon\psi$ 
is written as:

  $$ \pmatrix{m&\vec \sigma \cdot (\vec p+e \vec A)\cr
                   \vec \sigma \cdot (\vec p+e \vec A)&-m\cr}
     \pmatrix{\phi\cr \chi\cr } =\epsilon
      \pmatrix{\phi\cr \chi\cr }  \eqno (1)    $$
      
 and the system is solved for $\phi$ when $\epsilon >0$ and for $\chi$ when
      $\epsilon <0$ .
   \par   
  When the field becomes time dependent, the choice of the vector potential for
  the magnetic field becomes physically relevant, different choices yielding
  the same magnetic field give rise to different electric fields.
   So a choice is unavoidable: when the magnetic field is taken as
   constant in direction and varying in strength a cylindrical configuration, 
   defined by $\vec A=-\um \vec r \wedge \vec B$, will be studied. 
    Decomposing the variables into transverse and longitudinal
    components (the longitudinal direction is called $z$),
     the equation for $\phi $ is
    
 $$[p_z^2+p_T^2+ \uq e^2 r_T^2 B^2+eB(L_z+\sigma_z)]\phi=[\epsilon^2-m^2]\phi
 \eqno (2)$$

 The variables $p_z$ and $\sigma_z$ are diagonalized with eigenvalues $k$ and
 $s$ with the position 
 $$\phi={1\over {\sqrt Z}}\e^{\i kz}\phi_{k,s}\;,$$
 we obtain
 
 $$[p_T^2+\uq e^2 r_T^2 B^2+eB(L_z +s)]\phi_{k,s}
 =(\epsilon^2-k^2-m^2) \phi_{k,s}\;.\eqno (3)$$
 For future use the solution of the remaining equation is expressed in
 term of the harmonic-oscillator operators. Following ref.[8] one can define:
 
 $$ a_v=(p_v/b-\i bv)/\sqrt{2}\quad {\rm with} \quad v=x,y \quad b^2=\um e B$$

 $$ a_d=(a_x-\i a_y)/\sqrt{2} \;,\;a_g=(a_x+\i a_y)/\sqrt{2}\;,\;
 N_d=a_d^{\dagger}a_d\;,\;N_g=a_g^{\dagger}a_g \eqno (4)$$
  and express in this way the operators $p_T\;,\;r_T\;,\;L_z$. 
 The final result is :
 $$\epsilon^2=m^2+k^2+eB(2N_d+1+s)\;. \eqno (5)$$
 The operators $N_d\;,\;N_g$ are standard number operators, we respectively call 
 $n\;,\;\rho$ their eigenvalues and remark two known results relevant for the
 next developments:
  the quantum number $\rho$ has no influence on the value of the energy;
 the lowest Landau level, defined by $n=0$ and $s=-1$,
 has an energy independent of the strength of $B$. 
 \par
 The second order equation for $\chi$ is the same as the equation for $\phi$;
 the normalized solutions of the original equation are:
 $$ \psi_{+}={\cal N}\pmatrix{\phi\cr {\hat {\cal P}\over {w+m}}\phi\cr }\quad 
,\quad
\psi_{-}={\cal N}\pmatrix{-{\hat {\cal P}\over {w+m}}\chi\cr \chi\cr  }
\eqno (6)$$

    $$ \hat {\cal P}= \vec \sigma \cdot (\vec p+e \vec A)\quad ,\quad
    w=|\epsilon|\quad ,\quad {\cal N}=\sqrt{(w+m)/2w}\eqno (6')$$  
    
 The second quantized electronic field is decomposed in terms of these solutions
 $$\Psi=\sum_j c_j \psi_{+,j}+\sum_j d^{\dagger}_j \psi_{-,j} \quad      
\Psi^{\dagger}=\sum_j c^{\dagger}_j \psi^{\dagger}_{+,j}+
  d_j  \psi^{\dagger}_{-,j}\eqno (7)$$
  so the standard representation of the (second quantized) Hamiltonian is
  produced:
  $${\cal H}=\int d^3 r\Psi^{\dagger}(\vec r)
  [\vec \alpha \cdot (\vec p +e\vec A)+\beta m]\Psi (\vec r)=
  \sum_j w_j[c^{\dagger}_j c_j- d_j d^{\dagger}_j]\;.\eqno (8)$$
The index $j$ embodies all the relevant quantum numbers $j\equiv (k,n,\rho ,s)$,
 of course when they are read as positron quantum numbers, the sign of $k$ 
 and $s$ must be reversed. 
 \par
 Now the variation of the Hamiltonian with time is given writing \hfil\break
 $\vec B=B \vec \kappa$ and $B=B_o+\dot B t$ and defining
 $h={\partial H}/{\partial B}$ and
 $\alpha_{\pm}=\um(\alpha_x \pm \i \alpha_y)$. In terms of the 
 harmonic-oscillator operators, eq.(4), it results:
$$h=
 -\um e\vec \alpha \wedge \vec r \cdot \vec \kappa
 =\i {e\over {2b}}\bigl[-\alpha_{+} a_d+\alpha_{-} a_d^{\dagger}-
   \alpha_{-} a_g+\alpha_{+} a_g^{\dagger}\bigr]\;,\eqno (9)$$
 The derivative of the second quantized Hamiltonian is then:
 $$\eqalign{&{{\partial{\cal H}}\over{\partial B}}=\int d^3 r \Psi^{\dagger} h 
 \Psi=\cr
 &\sum_{j,j'}\bigl[c_j^{\dagger} c_{j'}
 \psi_{+,j}^{\dagger} h \psi_{+,j'} +c_j^{\dagger} d_{j'}^{\dagger}
 \psi_{+,j}^{\dagger} h \psi_{-,j'} +d_j c_{j'}
 \psi_{-,j}^{\dagger} h \psi_{+,j'} +
 d_j d_{j'}^{\dagger}
 \psi_{-,j}^{\dagger} h \psi_{-,j'}\bigr] \;.}\eqno (10)$$
 When we are interested in the transition from the vacuum to the
 electron-positron pair configuration only the second of the four addenda is
 interesting and will therefore be calculated explicitly; all the addenda 
 are important for higher order corrections.
 \par
 Before going on with the calculation of the matrix element, the following
 observation (ref. [8]), is useful: the mean value of $r_T^2$ is found to be
  $$<n,\rho| r_T^2 |n,\rho >=(n+\rho +1)/b^2$$
  In semiclassical states not having sharp
  eigenvalues the mean value becomes \footnote*{this procedure is similar to the
   treatment of
  a condensed Bose system, where one puts $a\approx a^{\dagger} \approx
  \sqrt{a^{\dagger}a}.$ }
  $$<\!<n,\rho| r_T^2 |n,\rho >\!>\approx (\sqrt{\rho}-\sqrt{n})^2/b^2\;.$$
  In both cases one observes that $n$ cannot grow too much because it costs
  energy eq.(5), while $\rho$ can grow without cost; so if the space at disposal 
of
  the system is large in most configurations we shall find
  $$<n,\rho| r_T^2 |n,\rho >\approx\rho/b^2 \;.$$
  It is, therefore, convenient to define $\rho=b^2 R^2$ and $R$ may be 
  interpreted as
  the mean position; the sum over $\rho$ which at the end occurs, is
  converted into an integration according to $\sum_{\rho}\approx b^2\int dR^2$.
   The region of integration will certainly be large, as
   compared with the microscopic lengths, so the terms 
   containing $R$ will be the most important and they will be calculated using
   the approximate representation $a_g\approx bR \e^{\i\theta}\;,\;
  a_g^{\dagger}\approx bR \e^{-\i\theta}$.
  The conclusion is that the dominant contribution for macroscopic extension of 
  the magnetic field may be extracted after reducing the expression of $h$ to
  the much simpler form
  $$h_o=
  \um \i e R\bigl(\e^{-\i \theta}\alpha_{+}-\e^{\i \theta}\alpha_{-}\bigr)\;.
  \eqno (11)$$
   \par
  The matrix element corresponding to the transition from the vacuum to a
  two-particle state is represented as
  $\int d^3 r\psi_{+,j}^{\dagger} h_o \psi_{-,j'}=
  \um \i e R{\cal N}_j{\cal N}_{j'}\,M$, two quantum 
  numbers are equal in $j$ and $j'$, they are $k$ and $\rho$.
   Using the relation:
$$ \hat {\cal P}= \sigma_z k+b\sqrt {2}[\sigma_+ a_d + \sigma_-a_d^{\dagger}]$$
 the expression for $M$ is found to be
 $$\eqalign {M=&\Bigl\{
  (\sigma_{+} \e^{-\i \theta}-\sigma_{-} \e^{\i \theta})-
  {1\over {(w_j+m)(w_{j'}+m)}}\times\cr
  &\bigl[-k^2(\sigma_{+} \e^{-\i \theta}-\sigma_{-} \e^{\i \theta})+\sqrt{2} kb 
  (\sigma_z a_d^{\dagger}\e^{-\i \theta}-\sigma_z a_d \e^{\i \theta}) +\cr
  &2b^2 (-\sigma_{+} a_d a_d\e^{\i \theta}+  
    \sigma_{-} a_d^{\dagger} a_d^{\dagger}\e^{-\i \theta})\bigr]\Bigr\}\;.}
    \eqno (12)$$
    
  From this form it appears that the two produced particles cannot
  have wholly equal quantum numbers, in particular it is impossible
  that both are produced in the ground state. There are eight relevant 
  matrix elements which are listed as follows.
  
  $$ \eqalign {T_1&=<-,n|M|+,n>=-\e^{\i\theta}-k^2\e^{\i\theta}
                  [(w_{+,n}+m)(w_{-,n}+m)]^{-1} \cr
               T_2&=<+,n|M|-,n>=\e^{-\i\theta}+k^2\e^{-i\theta}
                  [(w_{+,n}+m)(w_{-,n}+m)]^{-1} \cr
               T_3&=<+,n|M|+,n+1>=\sqrt {2} kb\e^{\i\theta}\sqrt {n+1}
                  [(w_{+,n}+m)(w_{+,n+1}+m)]^{-1} \cr
               T_4&=<+,n+1|M|+,n>=-\sqrt {2} kb\e^{-\i\theta}\sqrt {n+1}
                  [(w_{+,n}+m)(w_{+,n+1}+m)]^{-1} \cr
               T_5&=<-,n|M|-,n+1>=-\sqrt {2} kb\e^{\i\theta}\sqrt {n+1}
                  [(w_{-,n}+m)(w_{-,n+1}+m)]^{-1} \cr
               T_6&=<+,n+1|M|+,n>=\sqrt {2} kb\e^{-\i\theta}\sqrt {n+1}
                  [(w_{-,n}+m)(w_{-,n+1}+m)]^{-1} \cr   
               T_7&=<+,n|M|-,n+2>=2b^2\e^{\i\theta}\sqrt {(n+1)(n+2)}
                  [(w_{-,n+2}+m)(w_{+,n}+m)]^{-1} \cr
               T_8&=<-,n+2|M|+,n>=-2b^2\e^{-\i\theta}\sqrt {(n+1)(n+2)}
                  [(w_{+,n}+m)(w_{-,n+2}+m)]^{-1} }\eqno (13)$$  
  
  Further simplification is gained for strong magnetic fields, in fact in this 
  case the excited Landau levels are very far from the fundamental one so the 
  production of an electron-positron pair in their ground state would
  strongly be favoured, but it has just been seen that this transition is not
  allowed and at most one particle may be found in the ground state of its
  transverse motion. We may neglect the mass term in comparison with the 
  magnetic energy and write:
  $$ \eqalign{w_{+,n}&=\sqrt {4b^2 (n+1)+k^2}\; ,\;
  w_{-,n}=\sqrt {4b^2 n+k^2}\; {\rm if} \; n>0\cr
  w_{-,0}&\equiv w_k =\sqrt {m^2+k^2}\;.}\eqno (14)$$
 The dominant terms for large $b$ of the matrix elements are independent of $b$
 and they are explicitly:
    $$ \eqalign {\tilde T_1=&-\tilde T_2^{*}\approx -\e^{\i \theta}\cr          
                 \tilde T_3=&-\tilde T_4^{*}\to 0 \cr 
  \tilde T_5=&-\tilde T_6^{*}\approx -\e^{\i \theta} k/(w_k +m)\;{\rm if}\;n=0
           \quad \tilde T_5=-\tilde T_6^{*}\to 0\; {\rm if}\;n>0\cr
                 \tilde T_7=&-\tilde T_8^{*}\approx \um\e^{\i \theta} \;.}
                 \eqno (13')$$
  The same approximation is introduced into the normalization factors and for 
  them it results ${\cal N}\approx 1/\sqrt {2}$, but for 
  ${\cal N}_{-,0}=\sqrt{(w_k+m)/2w_k}$.
  The amplitude for the pair production is written in terms of the standard
  adiabatic expression [7], with the following specifications:
  only the transition from the vacuum to the two particle state is considered;
  the reduction of amplitude of the vacuum state is not taken into account; the
  initial instant of the process is not pushed to $-\infty$, but it is some 
  finite time, $t=0$, in which the field $B(0)=B_o$ is already large. 
  This is expressed writing the adiabatic projection coefficient
  [7] $\gamma_{j,j'}$ as:
  $$\gamma_{j,j'}(t)=-\dot B\int_o^t d\tau 
  {{<j,j'|\partial {\cal H}/\partial B|\,>}
     \over {\Delta E}} \gamma_o 
     \exp\bigl[-\i\int_o^{\tau}\Delta E dt' \bigr]\;:$$
     
  with $\Delta E=w_j +w_{j'} $, eq (14).
  \par
  From now on we refer to the matrix elements $T_1,\,T_2$, in the other cases
  slight modifications are required.
  \par
 It is convenient to separately treat the $n=0$ and the $n>0$ contributions.
  In this last case, in which the sum over all $n$ will be performed, we shall 
  set $n\approx n+1$ and in so doing 
  $$\int_o^{\tau}\Delta E dt'={2\over {3e\dot B n}}\bigl[ (2enB(t)+k^2)^{3/2}-
  (2enB_o+k^2)^{3/2}\bigr]\;;$$
  the second term gives rise to an irrelevant constant phase, in the calculation
  of $|\gamma_{j,j'}(t)|=\um eR \dot B |{\cal J}|$  the relevant integral is
  $${\cal J}=\int_o^t {{d\tau}\over{2\sqrt{2enB(\tau)+k^2}}}
  \exp\Bigl[-{{2\i}\over {3en\dot B}} (2enB(\tau)+k^2)^{3/2}\Bigr]\;.$$
 A convenient form for this expression is    
  $$\eqalign { {\cal J}=&{1\over {2en\dot B}}\bigl[f(S_o)\sqrt {2enB_o+k^2} +
     f(S(t) \sqrt{2enB(t)+k^2} )\bigr]\cr
   f(q)=&\int_1^{\infty}\exp [-\i q y^3]\,dy\cr
  S_o=& {2\over {3en\dot B}} \bigl(2en B_o +k^2\bigr)^{3/2}\cr 
  S(t)=&{2\over {3en\dot B}} \bigl(2en B(t) +k^2\bigr)^{3/2}\;.}
  \eqno (15)$$
  The parameters $S$ are very large in all the interesting situations. This is
  better appreciated if we insert the powers of $ c, \hbar$ that are needed to
  come back to standard units, for simplicity , when $k=0$  
   $$S={{4c\sqrt{2e/\hbar}}\over{3\dot B}}\sqrt{n}B^{3/2}$$
   then from the general asymptotic expansion of $f$,
   $$f(q)\simeq {1\over {3\i q}}\e^{-\i q}
   \sum_{\ell}\Bigl({2\over 3}\Bigr)_{\ell}\Bigl({\i\over q}\Bigr)^{\ell}$$
   the first term is sufficient \footnote
 *{it is possible to give a close expression for $f$, in terms of the irregular
   confluent hypergeometric function $U$ [9], although this is not very useful.}
    and we obtain:
   $${\cal J}=-\uq\i\bigl[\bigl(2en B_o +k^2\bigr)^{-1}\exp [\i S_o]-
      \bigl(2en B(t) +k^2\bigr)^{-1}\exp [\i S(t)]\bigr]\;.$$
  Moreover, since $S(t)$ is very large, the term containing
   $\cos[S_o-S(t)]$ which appears in $ |{\cal J}|^2$ is rapidly varying 
   around 0 and it will be dropped in the following.
   This us allows to write:
  $$ |{\cal J}|^2\approx {1\over{16 (2enB_o+k^2)^2}}+
  {1\over{16 (2enB(t)+k^2)^2}} \eqno(16)$$
  The sum over the longitudinal quantum number is performed with the usual
  transition to the continuum variable $\sum_k \to \udp Z\int dk$, the sum
  over the transverse quantum number remains discrete and the final result is
  $$ \sum _{k,n}|{\cal J}|^2\approx {Z\over 128} \zeta(3/2) 
 \Bigl[ {1\over {(2eB_o)^{3/2}}}+{1\over {(2eB(t))^{3/2}}}\Bigr]\;. \eqno (17)$$
   The symbol $\zeta$ represents the Riemann's z-function [9].
   \par
   For the estimate of the $n=0$ term the difference between the two addenda 
   $(+)$ and $(-)$ is important and we obtain:
   $$\int_o^{\tau}\Delta E dt'=k\tau +{1\over {3e\dot B n}}
   \bigl[ (2enB(t)+k^2)^{3/2}-(2enB_o+k^2)^{3/2}\bigr]=\Phi\;;$$
   The subsequent integral is more complicated:
   $${\cal J}_o=\int {{d\tau}\over {\dot \Phi}} \e^{-\i\Phi}\;,$$
     but has the same qualitative features as in the previous case, $\Phi $ is 
     large and we consider the asymptotic expansion:
    $${\cal J}_o={\i\over {\dot \Phi^2}} \e^{-\i\Phi}-
    {{2\ddot \Phi}\over {\dot\Phi^4}} \e^{-\i\Phi}+\cdots\;.$$
    It may be verified that ${\ddot \Phi}/{\dot\Phi^2}$ is of the order $1/S$.
    Going on as in the previous case, we keep only the first term in the
    expansion, we drop the rapidly oscillating term and we obtain:
    $$|{\cal J}_o|^2\approx\bigl[k+\sqrt{2eB_o+k^2}\bigr]^{-4}+
    \bigl[k+\sqrt{2eB(t)+k^2}\bigr]^{-4} \eqno (18)$$
    The sum over the longitudinal quantum number yields:
     $$ \sum _{k}|{\cal J}_o|^2\approx {2Z\over 15 \pi} 
 \Bigl[ {1\over {(2eB_o)^{3/2}}}+{1\over {(2eB(t))^{3/2}}}\Bigr]\;. \eqno (19)$$
  Now we must simply collect the terms, $i.e.$ the factor $\uq (eR\dot B)^2,$
  relating $|{\cal J}|^2$ to $|\gamma|^2$ and the integration $\um eB \int 
dR^2$,
  which performs the sum over the remaining quantum number $\rho$, the final
  result is:
  $$I(t)={1\over {16\sqrt{2}}}
  \Bigl[{1\over 256} \zeta (3/2)+{1\over {15\pi}}\Bigr]
  Z R_M^4 e^{3/2} \dot B^2 
  \Bigl[{{B(t)} \over {B_o^{3/2}}}+{1\over {\sqrt{B(t)}}}\Bigl]\;.\eqno (20)$$
  The production rate in a cylinder of height $Z$ and radius $R_M$, in a uniform
  and very strong magnetic field, varying linearly with time is given
  approximately by $dI/dt$.
   If we evaluate $dI/dt$ we obtain a cubic
   dependence on $\dot B$, built up by an original
   linear dependence of the transition amplitude, which is standard in the
   adiabatic processes and has to be squared, and by the dependence on $B$
   which is mainly related with the fact that the produced state extends over 
   a transverse region becoming
   smaller and smaller as the magnetic field grows.
   
   While the dependence on the longitudinal variable is trivial the dependence
   on the transverse one, $i.e.$ the fourth power, is less obvious and it may be
   understood if we remember that the configuration is not translationally
   invariant in the transverse dimension. In fact, although the magnetic field 
   is uniform, the electric field $\cal E$ grows linearly with $R$ and this puts
    an evident
   limitation to the configurations one can consider: in fact too large values 
   of $R$ such that ${\cal E} =\um\dot B R >B$ cannot fit in this treatment.
   It is pleasant to find a numerical confirmation of the intuitive idea that 
   the $n=0$ configuration must be
   important, in fact $|{\cal J}_o|>|{\cal J}|$.
   The rest of the factors strongly depend on the details of the chosen
   configuration. In particular there must be a further and not very large 
   factor which takes into account the other transition-matrix elements 
   $T_i\;,\;i>2$ .
   \vskip 1pc
   
   {\bf 3. Conclusions and outlook}
   \vskip 1pc
   A calculation of the pair production in a time-dependent magnetic field has
   been presented, in a simple situation where the already strong field is
   still growing in a fixed direction, $\dot B>0$. This process is suggested by 
the
   consideration of an imploding neutron star where the magnetic field is
   locally compressed [6].
    The choice of a very strong 
   field has a less dramatic effect on the production rate than one could have
    expected at first sight. 
    In fact, to the growth of the state density in $x$-space there
   corresponds a growth in the distance among the energy levels and therefore
    a decrease
   of density in momentum space. A stronger production rate could take place
   if both produced particles could be found in the ground
   state, but the
   dynamics does not allow this process. We may note, however, that as the
   distance of the Landau
   level grows there is an increasing number of longitudinal levels between two
   transverse levels, the convergence in the longitudinal momentum is always
   rapid, as $dk/k^4$, so the dependence on $k$ which has been neglected in the
   terms $T_i$ cannot play any important role.
   A more formal and general observation can be done: the total angular momentum 
   $\vec J=\int d^3r\, \vec r\wedge [\vec {\cal E}\wedge \vec B]$
   is zero in the considered volume and field configuration while an
   $e^{-}\,e^{+}$ pair in the lowest energy level has no orbital momentum and
   opposite spins, so the total angular momentum is 1, the transition is
   therefore impossible.
   \par
   Another interesting configuration could be given by a magnetic field which 
   changes its
   direction, like $e.g.$ that one could find around a rotating neutron star
   [5].
   The adiabatic formalism is certainly still useful, a first look at the 
   problem shows that in this case some difference in the treatment is needed 
   since the
   adiabatic states are, of course, time dependent, but the energy levels remain
   constant. Anyhow a more detailed investigation is required in order to obtain
   a definite answer.
   \par
   Much more complicated is the possibility that an intense magnetic field may
   act together with the QCD vacuum leading to an enhanced production of quarks
   and so, finally, of hadrons. Similar effects have in fact been proposed [10],
   but any quantitative estimate is very uncertain.
   
\vskip 1pc
{\bf Acknowledgments}
\vskip 1pc
This argument was suggested to me by prof. G. Barbiellini who, more in general,
called my attention to the relevance of strong field QED processes in
astrophysics.
\par
This work has been partially supported by the Italian Ministry of the University
and of Scientific and Technological Research by means of the {\it Fondi per la
Ricerca scientifica - Universit\`a di Trieste }.
\vfill
\eject
{\bf References}
\vskip 1pc
\item{1.}J.Schwinger, Phys. Rev. 82, 664 (1951) and Phys. Rev. 93, 615 (1954)
\item{2.}V.I Ritus: The Lagrangian function of an intense electromagnetic field
         A.I.Nikishov: The S-matrix of QED with pair creating external field
         in {\it Issues in intense field QED;} ed. V.L. Ginzburg - Nova
         scientia pub. N.Y. 1987
\item{3.}E. Brezin, C. Itzykson Phys. Rev D2, 1191 (1970)
\item{4.}P. Pengo {\it et al.}: Magnetic Birefringence of Vacuum: the PVLAS
experiment; W.T. Ni: Magnetic Birefringence of Vacuum: Q \& A Experiment.
         in {\it Frontier tests of QED and the physics of the vacuum} ed. E.
         Zavattini, D. Bakalov, C.Rizzo - Heron press Sofia 1998
\item{5.}C.Thompson, R.C.Duncan, Astrophys. J. 408, 194 (1993)
\item{6.}H. Hanami, Astrophys. J. 491, 687 (1997);
         K. Hurley {\it et al.} , Nature 397, 41 (1999)
\item{7.}A.B. Migdal, V. Krainov {\it Approximation methods in quantum
mechanics} Benjamin, New York 1969 p.81 ff.
\item{8.}C. Cohen-Tannoudji, B. Diu, F. Lalo\"e {\it M\'ecanique quantique - 
tome I} Hermann, Paris 1977 p.751 ff.

\item{9.}M.Abramowitz, I.A.Stegun  {\it Handbook of Mathematical functions}
         $\quad$ Dover, New York 1964 Ch.6
\item{10.}J. Rafelski: Electromagnetic Fields in QCD Vacuum         
         in {\it Frontier tests of QED and the physics of the vacuum} ed. E.
         Zavattini, D. Bakalov, C.Rizzo - Heron press Sofia 1998

\vfill
\eject 
\end
\bye